\documentstyle[graphicx]{asu2}

\def\softd{{\leavevmode\setbox1=\hbox{d}%
\hbox to 1.05\wd1{d\kern-0.4ex{\char039}\hss}}}
\def\softt{{\leavevmode\setbox1=\hbox{t}%
\hbox to \wd1{t\kern-0.6ex{\char039}\hss}}}
\def\softl{l\kern-0.45ex\raise0.1ex\hbox{'}\kern-0.10ex}
\def\softL{L\kern-0.8ex\raise0.1ex\hbox{'}\kern0.1ex}

\begin{document}

\setpage{13}
\title{Long Baseline Interferometry of Be Stars}
\subtitle{A Basic Introduction and First Results from MIDI/VLTI}
\author{O. Chesneau\inst{1,2}
\and
Th. Rivinius\inst{3}
}
\institute{
Max-Planck-Institut
f\"ur Astronomie, K\"onigstuhl 17, 67117, Heidelberg, Germany;\\
chesneau@mpia-hd.mpg.de
\and
Observatoire de la C\^{o}te d'Azur,
D\'{e}partement Gemini UMR 6203, Avenue Copernic, F-06130 Grasse,
France
\and
Landessternwarte K\"{o}nigstuhl, 69117 Heidelberg,
Germany;\\ triviniu@lsw.uni-heidelberg.de
}
\abstract{We give an introduction to interferometrical concepts and
their applicability to Be stars. The first part of the paper
concentrates on a short historic overview and basic principles of
two-beam interferometric observations. In the second part, the
VLTI/MIDI instrument is introduced and its first results on Be stars,
obtained on $\alpha$\,Ara and $\delta$\,Cen, are outlined. }
\maketitle

\section{Introduction}
In the recent past, optical interferometry has made the greatest
impact in the area stellar astrophysics, in particular the study of
nearby single stars. Be stars are hot stars that exhibit, or have
exhibited the so-called Be phenomenon, i.e.\ Balmer lines in emission
and infrared excess, interpreted as an equatorial disk around these
objects. Be stars are relatively frequent among the B-type objects, and
therefore, many bright and close Be stars are known.

These stars have been preferred targets for long baseline
interferometry since long, and the Be community has followed the
new developments of optical long baseline interferometry to study
the circumstellar environments of Be stars with great interest
(see also the recent review of Stee \& Gies 2004).

The first environment resolved was the one of $\gamma$~Cas. Thom et
al.\ (1986) used the I2T for this, and Mourard et al.\ (1989) saw
evidence for a rotating circumstellar environment by inspecting the
visibility across the line itself using the GI2T.

These results clearly demonstrated the potential of observations that
combine spectral and spatial resolution, but also that extensive
modeling is required to interpret measurements obtained with very
limited sampling of the $uv$-plane. The first model specialized for
this task was the one developed by Stee \& de Araujo (1994) and Stee
et al.\ (1995). Their model represents the environment of a Be star as
an axisymmetric structure, based on a latitude-dependant radiatively
driven wind. The model confirms that its free parameters can be
constrained by comparison of predicted line profiles and visibilities.

With a good range of baselines, the Mark~III instrument was able
to determine the geometrical aspect of seven Be stars, i.e.\ the
axial ratio of their elongated H$\alpha$ circumstellar emission
region (Quirrenbach et al., 1993a, 1994, 1997). The axial ratios
$r$ span a wide range, with $r < 0.5$ for $\phi$\,Per, $\psi$\,Per
and $\zeta$\,Tau, an intermediate ellipticity ($r=0.7$) for
$\gamma$\,Cas, and $r\sim1$ for $\eta$\,Tau and 48\,Per. In the
disk model for Be stars, this can easily be understood as an
inclination effect. The strong correlation of the minimum
inclination derived in this way with polarimetric estimates
supports the geometrically thin-disk hypothesis (Quirrenbach et
al.\ 1997).

The Mark~III was specifically designed to perform wide-angle
astrometry, but a variable baseline that could be configured from 3m
to 31.5m provided the flexibility needed for a variety of astronomical
research programs. Mark~III was the first interferometer having a full
computer control of the siderostats and delay lines which allowed
almost autonomous acquisition of stars and data taking. This
capability was an important factor for the calibration of instrumental
effects and for the scientific productivity of the instrument.

Among the stars investigated with Mark~III were also $\gamma$\,Cas and
$\zeta$\,Tau. In their disks asymmetric H$\alpha$ emission with
respect to the central object was later uncovered with the GI2T
instrument.  Using spectral Differential Interferometry (DI) it has
become possible to monitor such structures of Be disks during a long
time (several years) with great spatial resolution (Vakili et al.\
1998, B\'erio et al.\ 1999).

The field of optical and infrared (IR) interferometry has seen rapid
technical and scientific progress over the past decade and the
interferometric efficiency improves now dramatically in the era of the
Very Large Telescope Interferometer (VLTI). The design of ESO's VLTI,
which is the first large optical/IR telescope facility expressly
designed with aperture synthesis in mind, is of a hybrid type: There
are the four large 8.2 meter spatially fixed unit-telescopes, but
additionally there will be four auxiliary telescopes of smaller, i.e.\
1.8-meter aperture, which can be moved and set up at a large number of
locations. Three recombiners are attached to this structure: VINCI was
foreseen to be a K band test instrument but has provided such precise
visibility measurements that numerous outstanding science results have
been published from its observation. MIDI is the first direct
recombiner operating in N band in the world, which is described
extensively in the following. Finally, AMBER, currently being in
commissioning phase, is an impressive three-beam
spectro-interferometer operating in J, H and K bands with a spectral
resolution reaching 10\,000.

A challenge for the understanding of the Be star phenomenon is the
rapid increase of our knowledge of the central star itself. Be
stars are statistically rapid rotators and subject to a strong von
Zeipel effect (1924). In 2001, van Belle et al. (2001) observed
Altair (HD 187642, A7V) using two baselines of the Palomar Testbed
Interferometer (PTI). They calculated the apparent stellar angular
diameters from measured squared visibility amplitudes using the
uniform-disk model and found that the angular diameters change
with position angle. This was the first measurement of stellar
oblateness owing to rapid rotation. In parallel, the observable
consequences of this effect on the interferometric observation
have been extensively studied by Domiciano de Souza et al. (2002)
under the Roche approximation (uniform rotation and centrally
condensed star). Two effects are competing affecting the
interferometric signal: the geometrical distortion of the stellar
photosphere and the latitudinal dependence of the flux related to
the local gravity resulting from the von Zeipel theorem. The
measurements from the PTI were not sufficient to disentangle
between these two effects but recent observations using closure
phases\footnote{Closure phases are measured when at least three
telescopes are operating simultaneously.} from the NPOI
interferometer reported in Ohishi et al. (2004) have confirmed the
oblateness of the star and have evidenced a bright region
attributed to the pole. The observations of Altair from PTI, NPOI
and also from VLTI/VINCI cover now three spectral regions visible,
H and K band and an extensive modeling of the star has been
undertaken by A. Domiciano de Souza.

Altair is still a relatively cool and small star compared to the
Be stars and its gravity surface remains large, therefore larger
effects of the rotation are expected for Be stars. In 2003, the
large oblateness of the Be star Achernar ($\alpha$\,Eri) was
measured with VLTI/VINCI (Domiciano de Souza et al. 2003). The
measured oblateness of 1.56$\pm$0.05 based on equivalent Uniform
Disk apparently exceeds the maximum distortion allowed in the
Roche approximation and no models investigated by Domiciano de
Souza et al. (2002) could be satisfactorily fitted to the
observations. These observations open a new area for the study of
the Be phenomenon and VLTI/AMBER should take over this study and
expand rapidly the number of target observed.

Recently, VLTI/MIDI observed two Be stars, $\alpha$\,Ara (B3\,Ve)
and $\delta$\,Cen (B2\,IVne) from 8 to 13\,$\mu$m with baselines
reaching 100\,m but their circumstellar environment could not be
resolved. These observations are also reported in this paper.

\section{Basic Principles of Stellar Interferometry}
This section will review the basic principles of stellar
interferometry. More detailed discussions of optical interferometry
issues can be found in the reviews by Monnier (2003) and by
Quirrenbach (2001). In order to introduce the principles, we restrict
ourselves to the case of a single interferometric baseline, i.e.\ with
two telescopes only. We adopt the formalism of Domiciano de Souza et
al.\ (2002) and reproduce here the equations necessary for an
introduction to natural light interferometry.

\subsection{Basic principles}
Let us consider an astrophysical target located at the center of a
Cartesian coordinate system $(x, y, z)$. The system is oriented such
that the $y$ axis is defined as the North-South celestial orientation
and the $x$ axis points towards the observer.

Next, we define the sky-projected monochromatic brightness
distribution $I_{\lambda}(y,z)$, hereafter called "intensity map".
Interferometers measure the complex visibility, which is proportional
to the Fourier transform of $I_{\lambda}(y,z)$, which shall be denoted
$\widetilde{I}_{\lambda}(y,z)$.  The complex visibility in natural
light can then be written as:
\begin{eqnarray}\label{eq:V}
V(f_{y},f_{z},\lambda)& = & \left| V(f_{y},f_{z},\lambda)\right|
\mathrm{e}^{\mathrm{i}\phi(f_{y},f_{z},\lambda)}\\
 & = & \frac{\widetilde{I}_{\lambda}(f_{y},f_{z})}{
\widetilde{I}_{\lambda }(0,0)}
\end{eqnarray}
where $f_{y}$ and $f_{z}$ are the Fourier spatial frequencies
associated with the coordinates $y$ and $z$. These spatial frequencies
in long-baseline interferometry are given by $\vec{B}_{\rm
proj}/\lambda_{\rm eff}$, where $\lambda_{\rm eff}$ is the effective
wavelength of the spectral band considered and $\vec{B}_{\mathrm{\rm
proj}}$ is a vector representing the baseline of the interferometer
projected onto the sky. The vector $\vec{B}_{\mathrm{proj}}$ defines
the direction $s$, which forms an angle $\xi $ with the $y$ axis so
that
\begin{equation}\label{eq:Bproj}
\vec{B_{\rm proj}}=\left( B_{\rm proj}\cos \xi \right) \widehat{y}+\left(
B_{\rm proj}\sin \xi \right) \widehat{z}
\end{equation}
where $\widehat{y}$ and $\widehat{z}$ are unit vectors.  Note that
{\em large} structure in image-space results in {\em small} structure
in Fourier transformed, i.e.\ visibility-space.


\begin{figure*}
\fbox{\parbox{\textwidth}{
\centerline{\bf Visibility and phase as observational concepts}

\begin{center}
\parbox{0.4\textwidth}{Interference pattern of a monochromatic point-source}

\includegraphics[angle=270,width=0.4\textwidth]{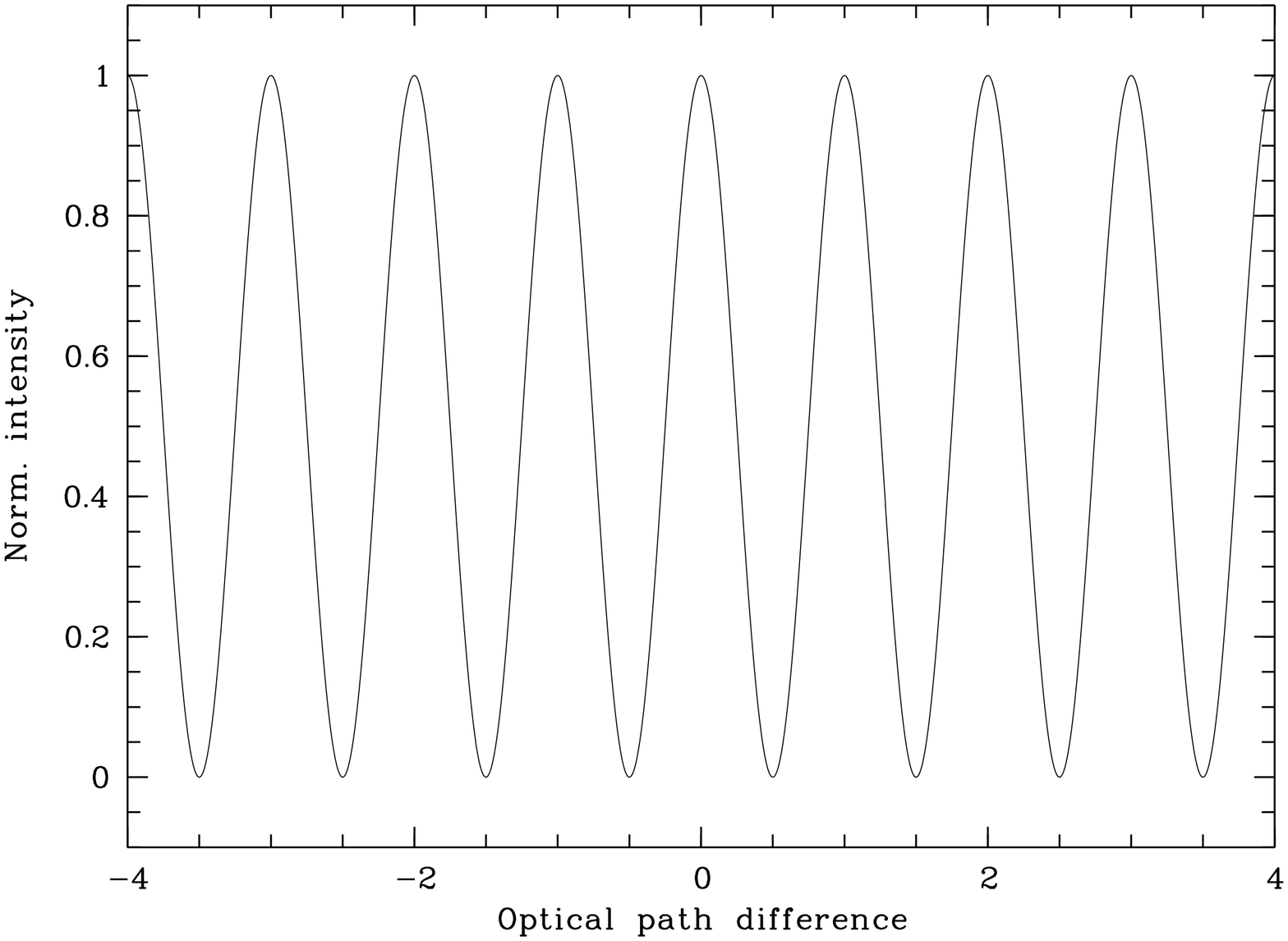}%

\parbox[t]{0.48\textwidth}{\begin{center}
\centerline{Polychromatic point-source}

\includegraphics[angle=270,width=0.4\textwidth]{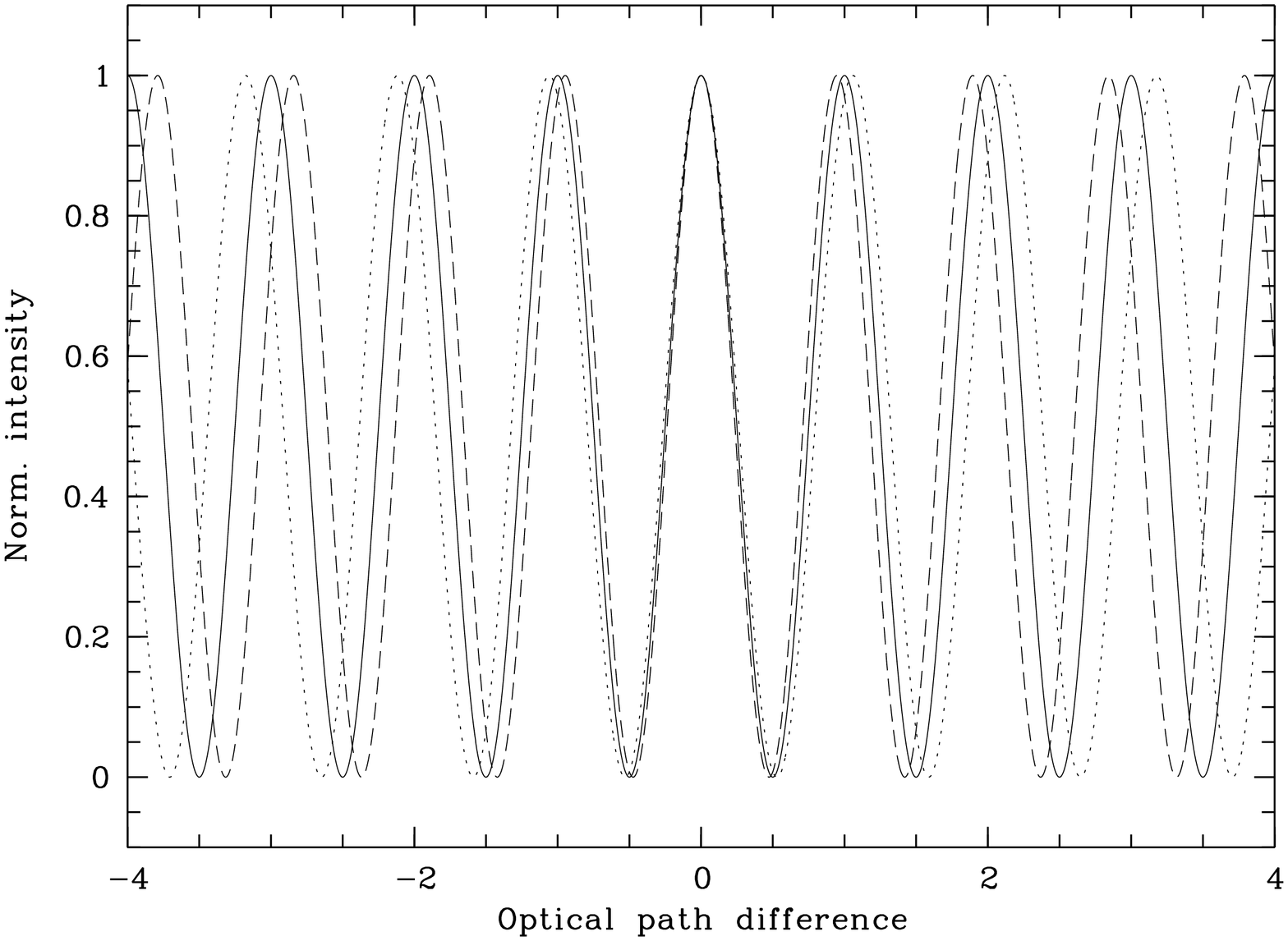}%

\includegraphics[angle=270,width=0.4\textwidth]{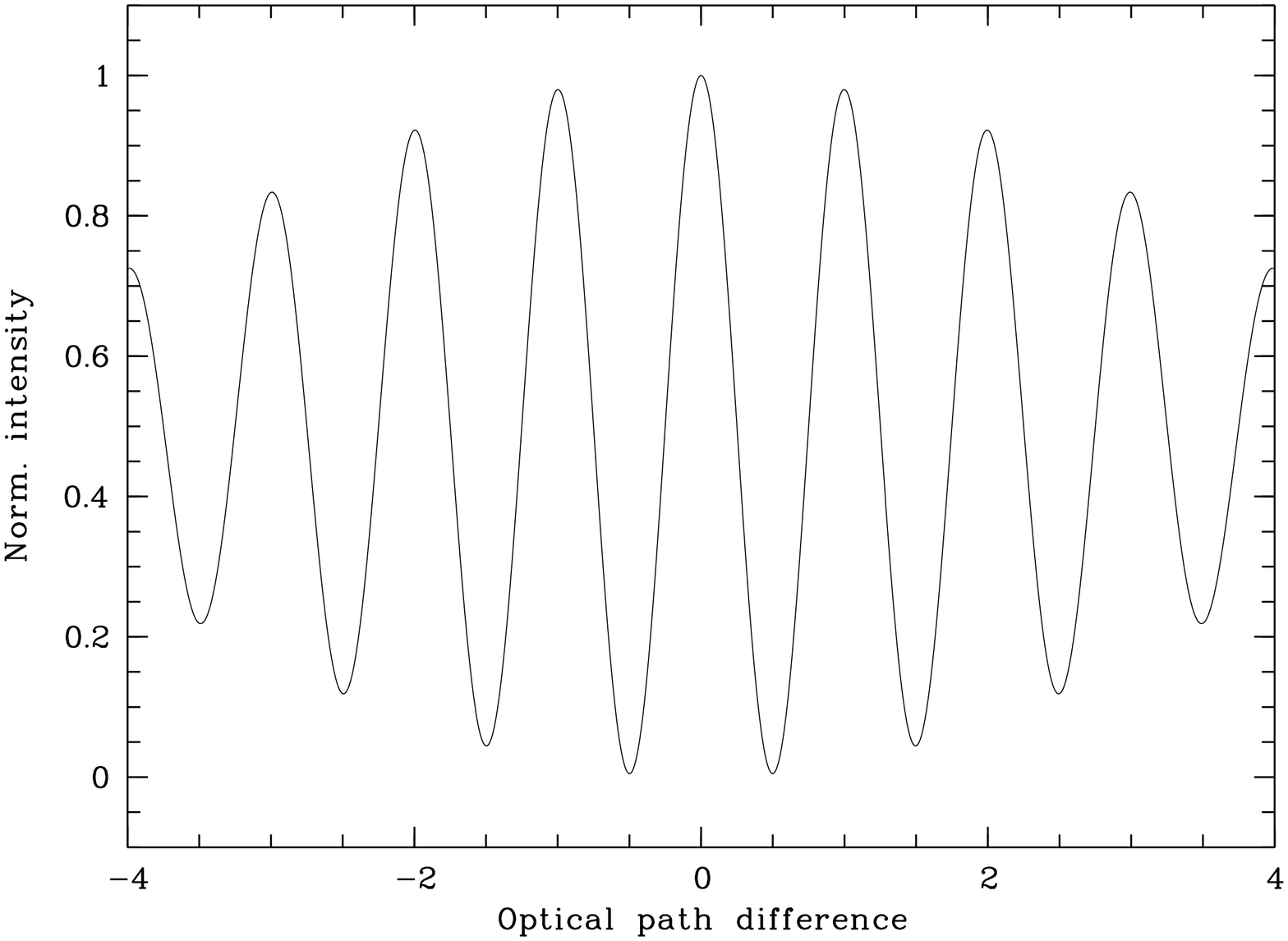}%
\vskip5mm

The interference patterns of the various colours add up to a spatial
modulation of the pattern. Note that the amplitude at OPD=0, the
``white-light fringe'' is still at maximum
\end{center}}\hspace*{0.04\textwidth}%
\parbox[t]{0.48\textwidth}{\begin{center}
\centerline{Monochromatic extended source}

\includegraphics[angle=270,width=0.4\textwidth]{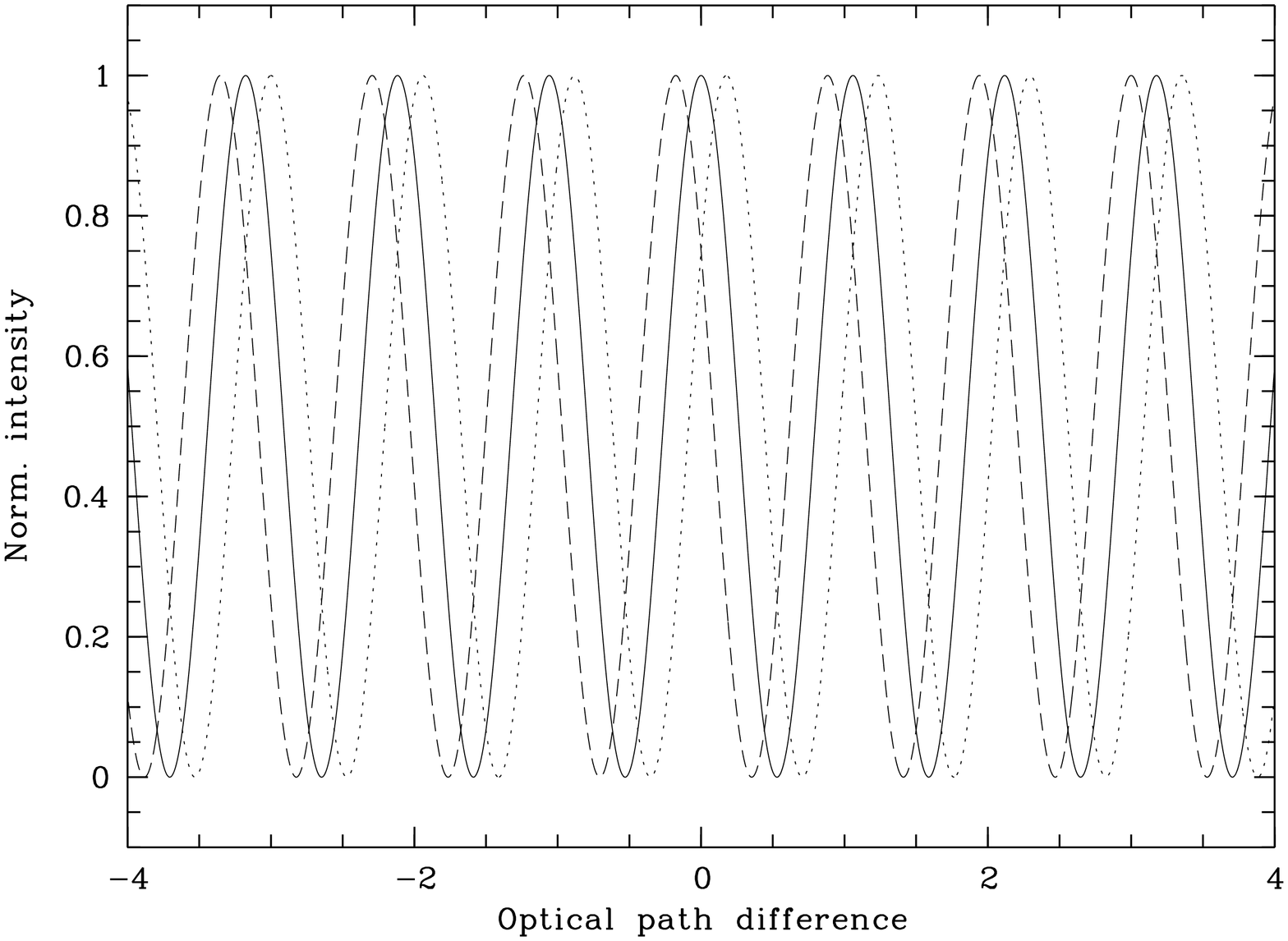}%

\includegraphics[angle=270,width=0.4\textwidth]{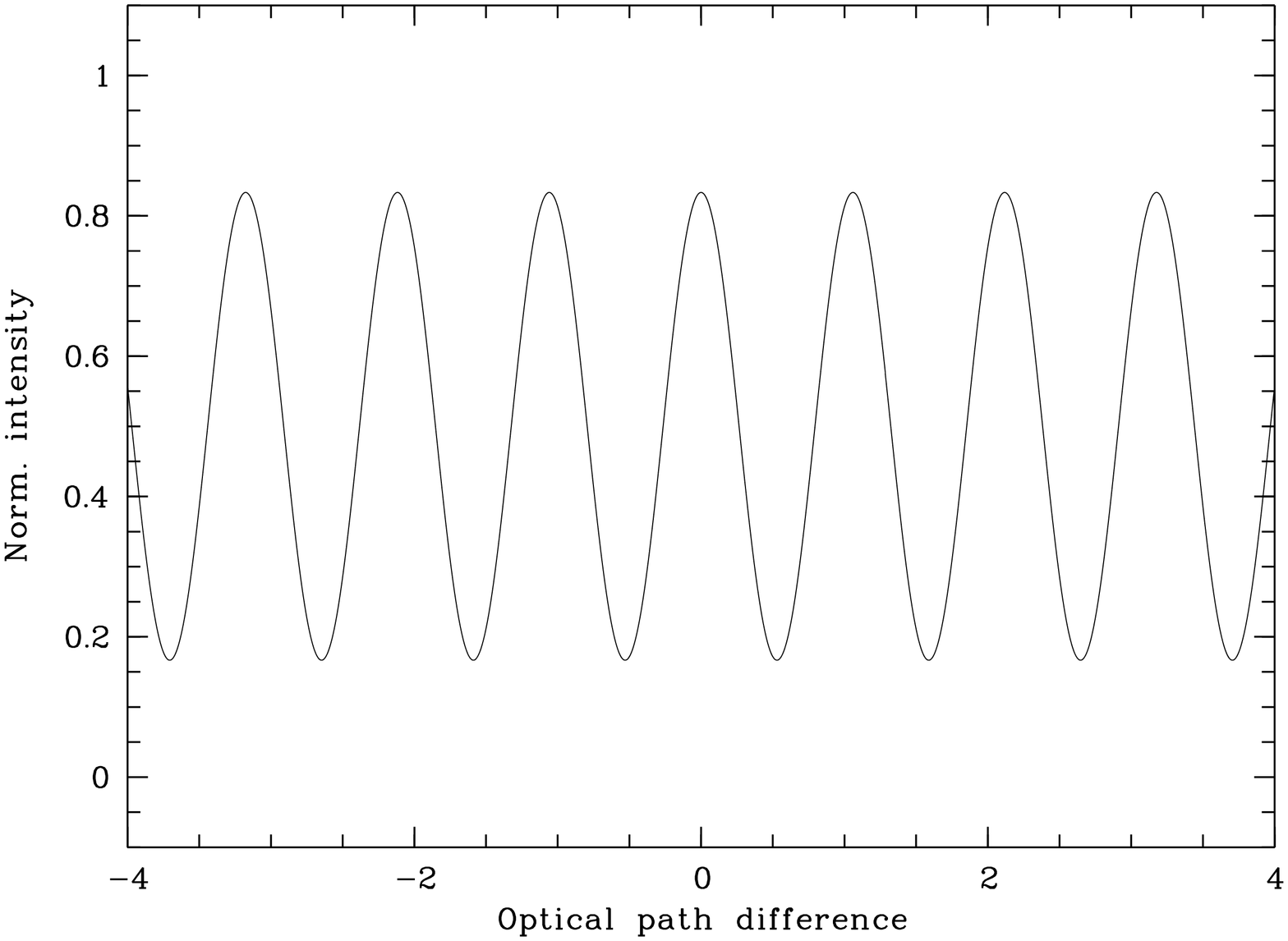}%
\vskip5mm

The interference patterns from the local points add up to a pattern of
reduced amplitude, independent of the optical path difference (OPD),
however.
\end{center}}%

\parbox[t]{\textwidth}{Both principles combined give the fringe
patterns as seen by interferometry. The ``visibility'', holding the
spatial extend of the investigated source is quantifying the
amplitude, as in the right column. Suppose a source observed at two
different wavelengths, one can obtain {\em relative} positional
information by measuring whether the OPD for maximal positive
interference has shifted. This concept is called ``interferometric
phase''}

\end{center}

}}
\end{figure*}

\begin{figure*}
\begin{center}
\includegraphics[width=0.83\textwidth]{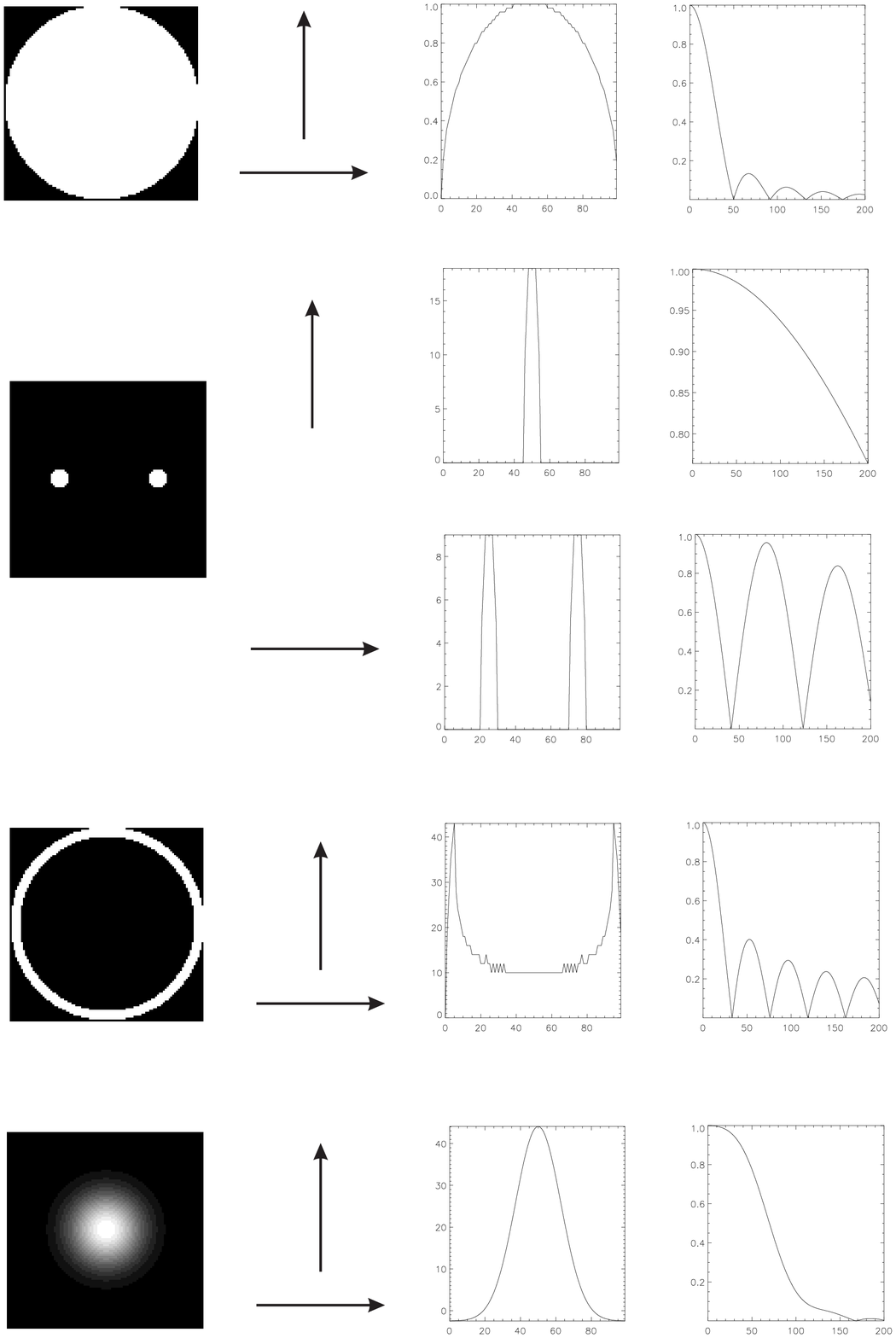}%
\end{center}
\caption{\label{fig1} In this figure typical examples of intensity
maps (left) are shown. From top to bottom these are a uniform disk, a
resolved binary, a ring and a Gaussian distribution. The models are
``observed'' with a horizontal ad vertical baseline. The 1-D flux
intensity along the baseline is shown in the middle and the
corresponding visibility is displayed on the right. All units are
arbitrary. A further description of visibility curves is given in the
text. }
\end{figure*}

We consider linear cuts along the Fourier plane corresponding to a
given baseline direction $\widehat{s}$. Then we can define the new
spatial frequency coordinates ($u,v$) for which
$\vec{B}_{\mathrm{proj}}$ is parallel to the unit vector
$\widehat{u}$. In that case the line integral (or strip intensity)
of $I_{\lambda }(s,p)$ over $p$ for a given $\xi$ can be written
as:
\begin{equation}\label{eq:FTline}
\widetilde{I}_{\lambda,\xi}(u)= \int I_{\lambda,\xi}(s)
\mathrm{e}^{-\mathrm{i}2\pi su} \mathrm{d}s
\end{equation}
The \textit{complex visibility} is given by:
\begin{equation}\label{eq:Vline}
V_\xi(u,\lambda)=\left| V_\xi(u,\lambda)\right|
\mathrm{e}^{\mathrm{i}\phi_\xi(u,\lambda)}=\frac{\widetilde{I}_{\lambda,\xi}(u)}{
\widetilde{I}_{\lambda,\xi}(0)}
\end{equation}
By varying the spatial frequency (meaning the baseline length and/or
wavelength), we obtain the so-called visibility curve. Eqs.\
\ref{eq:FTline} and \ref{eq:Vline} say that the interferometric
information along $\vec{B}_{\mathrm{proj}}$ is identical to the
one-dimensional Fourier transform of the curve resulting from the
integration of the brightness distribution in the direction
perpendicular ($\widehat{p}$) to this baseline. The
interferometric observable, called visibility is directly related
to the fringe contrast.

The visibility can be observed either as the fringe contrast in {\bf
an image plane} (as with AMBER) or by modulating the internal delay
and detecting the resulting temporal variations of the intensity in a
pupil plane (as done with VINCI or MIDI).

It should be stressed that the signal-to-noise ratio (SNR) of
interferometric observables depends not only on the photon count $N$,
but on $NV^2$ for the photon-noise limited regime (optical) and on
$NV$ for the background-limited regime (MIDI). Indeed, the
interferometer is not sensitive to the total flux from the source but
to the {\it correlated} one.

In Fig.~1 we show several examples of intensity maps and their
corresponding visibility curves, depending on the baseline
orientation. All models, except the binary, are symmetrical which
means that the interferometer will provide the same visibility for a
particular baseline length (projected onto the sky), whatever its
direction. The second model shows the visibility curves from a binary
system, consisting of two stars of the same diameter. In the case
where the baseline is perpendicular to the binary's position angle,
the interferometer is unable to distinguish the two components and
sees a visibility signal close to the one provided by a single uniform
disk. When the baseline is aligned with the binary's PA, the binary
signature is superimposed to the signature from the individual
components.

The uniform ring structure is an interesting intermediate
situation between the uniform disk and a binary. The
interferometer sees mainly a binary structure and owing to the
symmetry of the source, this signal is the same whatever the
baseline direction. This example could appear quite artificial, yet
it reflects a geometry which can be encountered frequently in the
inner rims of a Young Stellar Objects and even more evolved star's disks.

Finally, we show the example of an object exhibiting a Gaussian
distribution of light, like approximating a circumstellar environment
with outwards decreasing emission. The Fourier transform of a Gaussian
being a Gaussian too, the visibility curve will not show the
characteristic lobes which are seen in the other curves. It must be
stressed out that these lobes are the consequence in the Fourier
domain of a discontinuity of the light distribution in the image
plane. For instance, a limb-darkened disk will exhibit a visibility
curve with the lobes attenuated compared to the uniform disk case. One
can, in fact, note a small lobe in the Gaussian visibility curve in
Fig.~1. This is the consequence of the numerical truncation of the
Gaussian, generating a discontinuity at the limit of the chosen
Field-of-View.

Of course, for typical main sequence stars, this implies long
baselines and it must be stressed that at this stage the contrast of
the fringes is so low that this experiment is very demanding in terms
of signal-to-noise ratio, i.e.\ in terms of target flux.

\subsection{Differential Interferometry (DI)}
Differential Interferometry (DI) uses the high angular resolution
interferometric capabilities in a dispersed mode in order to compare
the spatial properties of an object at different wavelengths (Vakili
et al. 1994, 1997, 1998). This technique offers obvious advantages:
over few nanometers, the (differential) atmospherical turbulence
effects are negligible, and the differential sensitivity can be much
better than expected by classical techniques. In interferometry, an
unresolved source is required for calibration, and such an object can
be difficult to find. In particular cases, such as Be stars, the
continuum can be regarded as unresolved, whereas emission lines
emitted by an extended circumstellar environment are angularly
resolved. Moreover, the continuum can be considered as a {\em
polarization} reference in the Zeeman effect context.

The phase of the visibility is generally lost due to the blurring
effect of the atmosphere, but DI can retrieve a {\em relative}
spectral phase. Some algorithms can compare the properties of the
fringes studied in a (broad, i.e.\ continuum) reference spectral
channel at wavelength $\lambda_r$ with the ones in a (narrow, i.e.\
spectral line) science channel at wavelength $\lambda_s$. This can be
performed by means of cross-correlation of a broad continuum channel
with a series of narrow channels across the emission line as
encountered in Be stars. These steps are then repeated with small
wavelength shifts for both channels from the blue to the red wing of
the line, starting and finishing in the continuum next to the line on
both sides.  Since the signal to noise ratio in the cross-correlations
depends on the geometric mean of the number of photons in channel $r$
and channel $s$, the interferometric signal can be safely estimated in
the narrow channel even if the flux or the fringe visibility are very
small in this channel.

The accuracy of the phase determination, which allows to measure a
position can be better than the actual resolution of the
interferometer, but again the super-resolution power of
cross-spectral density methods apply, as long as the star is partially
resolved (Chelli \& Petrov 1995). For instance, a positive (negative)
relative phase indicates the position at the north (south) of the
central star if the baseline is oriented North-South. The relative
phase shift between to spectral channels is related to the photocenter
of the object by the following first order equation:

$$ \phi(\vec{u}, \lambda_r, \lambda_s)=-2 \pi
\vec{u}.[\vec{\epsilon}(\lambda_s)-\vec{\epsilon}(\lambda_r)]$$

Using the continuum fringes as reference for the phase one can,
for instance, determine whether the light distribution in a
spectral line is centered on this continuum, as any departure from
symmetry, e.g.\ due to localized circumstellar emission or a
magnetic field, causes a spectral phase effect. While the
visibility is a quadratic estimator, the phase sensitivity depends
linearly on the photon count.

Increasing theoretical predictions of photocenter positions
through emission or absorption lines are available concerning the
environment of stars (Stee et al. 1996, Dessart \& Chesneau 2002)
or the study of the underlying photosphere (Domiciano et al. 2002,
2004).

\section{MIDI Interferometer}
MIDI is the first 10~$\mu$m interferometric instrument worldwide using
the full available atmospheric transmission window. Due to the MIR
radiation of the environment and the optical setup itself, most of the
instruments optics is inside a dewar and is cooled to cryogenic
temperatures. The incoming afocal VLTI beams are combined on the
surface of a 50:50 beam splitter, which is the heart of the
instrument. Spectral information from 8~$\mu$m to 13.5~$\mu$m can be
obtained by spectrally dispersing the image using a prism for low
(R=30), or a grism for intermediate (R=270) spectral resolution. For
source acquisition a field of 3" is available.  This small area on the
sky represents about 10 times the Airy disk of a single UT
telescope. This field is useful especially in case of extended objects
like the S\,Doradus variable (also called LBV) $\eta$ Car (Chesneau et
al. 2004a) or some Herbig stars (like HD\,100\,546, Leinert et
al. 2004).

MIDI measures the degree of coherence between the interfering
beams (i.e. the object visibility) by artificially stepping the
optical path difference between the two input beams rapidly, using
its internal delay lines. The result is an intensity signal
modulated with time from which the fringe amplitude can be
determined. The total (uncorrelated) flux is determined separately
by chopping between the object and an empty region of the sky, and
determining the source flux by subtraction. In this mode MIDI is
working like an usual mid-infrared camera. An example of a
resulting image in case of the observation of $\eta$\,Carinae is
shown in Chesneau et al.  (2004a) which demonstrates the excellent
imaging capabilities of the MIDI/VLTI infrastructure, even if it
sends the light via 31 mirrors and 5 transmissive elements until
it reaches the detector.

Observing with an interferometer requires accurate preparation.
Useful tools for that are simulation programs like ASPRO (by the
Jean-Mariotti Center, Fr.), SIMVLTI (by the MPIA Heidelberg, Ge.)
or ESO's VLTI visibility calculator. Those software packages make
it possible to get an idea of the expected visibility values for
given parameter setups. For further reference, the reader should
also consult the MIDI web page at
ESO\footnote{http://www.eso.org/instruments/midi/}.

When planning observations with MIDI, a few constraints have to be
kept in mind. Of course, the object should be bright enough in the
mid-IR to be measured with MIDI. However, for self-fringe tracking,
the source not only has to be bright enough in total, but there must
be sufficient flux concentrated in a very compact
($<0.1^{\prime\prime}$) central region, to which then the
interferometric measurements will refer. Also, the {\em visual brightness}
should be at least 16 mag, in order to allow the operation of the
MACAO tip-tilt and adaptive optics system.

In addition, one has to consider that interferometry with two
telescopes of the VLTI in a reasonable time of several hours will
provide only few measured points of visibility, i.e.\ only a few points
where the Fourier transform of the object image is determined. The
scientific programme has to be checked before whether its main
questions can be answered on this basis (e.g.\ to determine the
diameter of a star one does not need to construct an image of its
surface).

\section{Be stars observed by MIDI}
\subsection{\boldmath$\alpha$ Ara\unboldmath}
The first VLTI/MIDI observations of the Be star $\alpha$~Ara show a
nearly unresolved circumstellar disk in the N band (Chesneau et
al. 2004b). $\alpha$~Ara (HD\,158\,427, B3\,Ve) is one of the closest Be
star with an estimated distance of 74pc$\pm$6pc, based on the
Hipparcos parallax, and color excesses E(V-L) and E(V-12 ~$\mu$m)
among the highest of its class. The interferometric measurements made
use of the UT1-UT3 projected baselines of 102~m and 74~m, at two
position angles of 7$^\circ$ and 55$^\circ$, respectively. The object
is mostly unresolved, putting an upper limit to the disk size in the N
band of the order of $\phi_{\rm max}=4$\,mas, i.e.\ 14 $R_\star$ at
74~pc and assuming $R_\star=4.8{\rm R_\odot}$, based on the spectral
type.

On the other hand, the density of the disk is large enough to produce
strong Balmer emission lines. The SIMECA code developed by Stee (1995)
and Stee \& Bittar (2001) has been used for the interpretation.
Optical spectra from the {\sc Heros} instrument, taken 1999, when the
instrument was attached to the ESO-50cm telescope, and infrared ones
from the 1.6m Brazilian telescope have been used together with the
MIDI visibilities to constrain the system parameters.  In fact, these
two observations, spectroscopy vs.\ interferometry, put complementary
constraints on the density and geometry of the $\alpha$~Ara
circumstellar disk.  It was not possible to find model parameters that
at the same time are able to reproduce the observed spectral
properties, both in the emission lines and in the continuum
flux-excess, and the interferometric null-result, meaning that the
disk must be smaller than 4\,mas.

However, the Hydrogen recombination line profiles of $\alpha$~Ara
exhibit (quasi?)-periodic short-term variations of the
violet-to-red peak heights of the emission profiles
($V/R$-variability, see Mennickent \& Vogt, 1991 and this study).
The radial velocity of the emission component of the Balmer lines
changes in a cyclic way as well in these lines. This may point to
a possible truncation of the disk by a putative companion, that
could explain the interferometric observations.

Using the NPOI interferometer, Tycner et al. (2004) have
recently studied the disk geometry of the Be star $\zeta$~Tau,
which is also a well-investigated spectroscopic binary
(P$\sim$133d, K$\sim$10\,km\,s$^{-1}$). They measured the disk
extension quite accurately to be well within the Roche radius.
This suggests also that this disk may be truncated.

\vspace{5mm}
\subsection{\boldmath$\delta$\,Cen\unboldmath}
$\delta$\,Cen (HD\,105\,435, B2\,IVe, F$_{12\mu m}$=15.85\,Jy),
situated at about 120\,pc, is one of the very few Be stars which
has been detected at centimeter wavelengths (Clark et al.\ 1998)
and also the only star for which significant flux has been
measured at 100$\mu$m (Dachs et al. 1988). These two measurements
suggest an extended disk, contrary to the case of $\alpha$~Ara.

However, recent VLTI/MIDI observations of $\delta$\,Cen during
Science Demonstration Time (programme conducted by D.\ Baade) with
a baseline of 91\,m have not been able to resolve the disk of this
object as well. It must be stressed out that these observations
have been conducted under much better atmospheric conditions than
those for $\alpha$\,Ara, leading to a well constrained upper limit
of 4$\pm$0.5\,mas for the equivalent Uniform Disk diameter.

This is roughly the same size as was determined for other Be stars
using interferometry in the wavelength region of H$\alpha$. Note that
for these H$\alpha$ observations the baseline was much less than the
one used here, about 40\,m vs.\ 100\,m. That both datasets still come
up with the same angular resolution is due to the scaling of the
spatial frequency with the effective wavelength introduced in
Sect.~2.1.

Whether or not a Be star disc should be resolved in the near IR
depends on the model one adopts for such a disk. Based on modelling of
the Balmer-line emission, at least one model does predict a resolved
disk, while others don't (see Chesneau et al., 2005, for a detailed
discussion). In this sense, even null-results provide important
constraints to our understanding of Be star disks.

\section{Conclusions}
Although being null-results these observations, as others before, have
shown the potential discriminating power of interferometric
observations for the current open questions of Be star research.  

Long baseline interferometry is now able to provide a complete set of
observations from the visible to the thermal infrared at high angular
and spectral resolution, opening a new area for the study of the Be
phenomenon. In particular, this technique is now able to study the
complex interplay between fast rotator distorted photospheres,
affected by the von Zeipel effect and their direct surroundings by
means of spectrally resolved NIR observations and MIR ones. The
Guaranteed Time Document of the VLTI/\-AMBER
\footnote{available at http://www-laog.obs.ujf-grenoble.fr/amber/}
gives a good idea of the possibilities opened by this new instrument.

The first VLTI/MIDI observations of Be stars have demonstrated the
need to use long baseline at these wavelengths in order to resolve the
disk of even the few closest (and brightest) Be stars. The VLTI 1.8\,m
Auxiliary Telescopes (ATs) AT1 and AT2 are currently being
commissioned at Paranal observatory and should be able to observe
their first fringes in mid-2005. The ATs are movable telescopes which
can project onto the sky a baseline from 8\, to 200\,m. Such long
baselines should be perfectly suited for MIDI to study the inner disk
of Be stars, and for AMBER to observe the star itself, whereas the
shorter ones would allow AMBER to study the close environment from the
photosphere to several stellar radii.

\vspace{15mm}

\references

\ritem B\'{e}rio, P., Stee, Ph., Vakili, F., et al., 1999, A\&A,
345, 203

\ritem Chelli, A., Petrov, R.G., 1995, A\&A Sup. Ser. 109, 389

\ritem Chesneau, O., Min, M., Herbst, T. et al.  2004a, A\&A,
submitted

\ritem Chesneau, O., Meilland, A., Stee, Ph et al. , 2004b, A\&A,
submitted

\ritem Clark J.S., Steele I.A., Fender R.P., 1998, MNRAS, 299,
1119

\ritem Dachs, J., Engels, D. and Kiehlin, R. 1988, A\&A, 194, 167

\ritem Dessart, L. and Chesneau, O., 2002, A\&A, 395, 209

\ritem Domiciano de Souza, A., Vakili, F., Jankov, S., 2002, A\&A,
393, 345

\ritem Domiciano de Souza, Kervella, P., Jankov, S. et al., 2003,
A\&A, 407, L47

\ritem Domiciano de Souza, A., Zorec, J., Jankov, S. et al. 2004,
A\&A, 418, 781

\ritem Leinert, Ch., van Boekel, R., Waters, L.B.F.M. 2004, A\&A,
423, 537

\ritem Monnier, J. Reports on Progress in Physics, 2003, Vol. 66,
789-857

\ritem Mourard, D., Bosc, I., Labeyrie, A. et al. 1989, Nature,
342, 520

\ritem Ohishi, N., Nordgren, T.E. and Hutter, D.J. 2004, ApJ, 612,
463

\ritem Quirrenbach, A., Bjorkman, K.S., Bjorkman, J.E. et al.,
1997, ApJ, 479, 477

\ritem Quirrenbach, A., 2001, Annual Review of Astronomy and
Astrophysics, Vol. 39, 353-401.

\ritem Quirrenbach, A., Buscher, D.F., Mozurkewich, D. 1994, A\&A,
283, L13

\ritem Quirrenbach, A., Hummel, C.A., Buscher, D.F. et al., 1993,
ApJ, 416, L25

\ritem  Stee, Ph.,  de Araujo, F.X., 1994, A\&A, 292, 221

\ritem Stee, P., de Araujo, F. X., Vakili, F. 1995, A\&A, 300, 219

\ritem Stee, P. 1996, A\&A, 311, 945

\ritem Stee, P. \& Gies, D. 2005, ASP Conf. Ser., in press

\ritem Thom, C., Granes, P., Vakili, F., 1986, A\&A, 165, L13

\ritem  Tycner, Ch., Hajian, A.R., Armstrong, J.T. et al., 2004,
AJ, 127, 1194

\ritem  Vakili, F., Mourard, D., Stee, Ph., et al. 1998, A\&A,
335, 261

\ritem Vakili, F., Mourard, D., Bonneau, D., 1997, A\&A, 323, 183

\ritem Vakili, F., Bonneau, D., Lawson, P.R., 1994, SPIE, 2200,
216

\ritem  van Belle, G.T., Ciardi, D.R., Thompson, R.R. et al. 2001,
ApJ, 559, 1155

\ritem  von Zeipel, H. 1924, MNRAS, 84, 665


\end{document}